\documentclass[%
 aip,
 mph,%
 amsmath,amssymb,
 superscriptaddress,
 floatfix,
preprint, 
eqsecnum,%
numerical,%
]{revtex4-1}

\usepackage{graphicx}
\DeclareGraphicsExtensions{.eps,.pdf,.png,.jpg}
\usepackage[caption=false
]{subfig}
\usepackage{epsfig}
\usepackage{epstopdf}
\usepackage{lipsum}
\usepackage{wrapfig}
\usepackage{makecell}
\usepackage{tabularx}
\usepackage{multirow}
\usepackage{soul}
\usepackage{booktabs}
\usepackage{lmodern}
\usepackage{longtable}
\usepackage{amsmath}
\usepackage[OT2,T1]{fontenc}
\usepackage[bookmarks=false]{hyperref}
\usepackage{color}
\usepackage{float}
\hypersetup{pdfauthor={},pdfborder=0 0 0,colorlinks=true,citecolor=blue,linkcolor=blue,urlcolor=blue,}

\DeclareSymbolFont{cyrletters}{OT2}{wncyr}{m}{n}
\DeclareMathSymbol{\comb}{\mathalpha}{cyrletters}{"58}

\newcommand{\ben}{\begin{eqnarray}\displaystyle}
\newcommand{\een}{\end{eqnarray}}

\begin{document}

\title{Scatter correction for photon-counting detector based CBCT imaging}

\author{Xin Zhang}
\affiliation{Research Center for Advanced Detection Materials and Medical Imaging Devices, Shenzhen Institute of Advanced Technology, Chinese Academy of Sciences, Shenzhen, Guangdong 518055, China} 
\affiliation{University of Chinese Academy of Sciences, Beijing 100049, China}

\author{Ting Su}
\affiliation{Research Center for Advanced Detection Materials and Medical Imaging Devices, Shenzhen Institute of Advanced Technology, Chinese Academy of Sciences, Shenzhen, Guangdong 518055, China}

\author{Jiongtao Zhu}
\affiliation{Research Center for Advanced Detection Materials and Medical Imaging Devices, Shenzhen Institute of Advanced Technology, Chinese Academy of Sciences, Shenzhen, Guangdong 518055, China} 

\author{Hairong Zheng}
\affiliation{Paul C Lauterbur Research Center for Biomedical Imaging, Shenzhen Institute of Advanced Technology, Chinese Academy of Sciences, Shenzhen, Guangdong 518055, China}
\affiliation{Key Laboratory of Biomedical Imaging Science and System}

\author{Dong Liang}
\affiliation{Paul C Lauterbur Research Center for Biomedical Imaging, Shenzhen Institute of Advanced Technology, Chinese Academy of Sciences, Shenzhen, Guangdong 518055, China}
\affiliation{Key Laboratory of Biomedical Imaging Science and System}
\affiliation{Research Center for Medical Artificial Intelligence, Shenzhen Institute of Advanced Technology, Chinese Academy of Sciences, Shenzhen, Guangdong 518055, China}
\author{Yongshuai Ge}
 \thanks{Scientific correspondence should be addressed to Yongshuai Ge (ys.ge@siat.ac.cn).}
\affiliation{Research Center for Advanced Detection Materials and Medical Imaging Devices, Shenzhen Institute of Advanced Technology, Chinese Academy of Sciences, Shenzhen, Guangdong 518055, China}
\affiliation{Paul C Lauterbur Research Center for Biomedical Imaging, Shenzhen Institute of Advanced Technology, Chinese Academy of Sciences, Shenzhen, Guangdong 518055, China}
\affiliation{Key Laboratory of Biomedical Imaging Science and System}
\affiliation{National Innovation Center for Advanced Medical Devices, Shenzhen, Guangdong 518131, China.}

\date{\today}

\begin{abstract}

{\bf Objective:} The aim of this study is to validate the effectiveness of an energy-modulated scatter correction method in suppressing scatter in photon-counting detector (PCD)-based cone beam CT (CBCT) imaging.\\
{\bf Approach:} The scatter correction method, named e-Grid, which was initially applied to dual-layer flat-panel detector (DLFPD)-based CBCT imaging, was tested for its performance in PCD-CBCT imaging. Benchtop PCD-CBCT imaging experiments were conducted to verify the effectiveness of the e-Grid method. Additionally, quantitative metrics were measured from these experimental results. \\
{\bf Main results:} It was found that the use of the e-Grid method could significantly eliminate cupping artifacts caused by Compton scatter in PCD-CBCT imaging. Meanwhile, its effectiveness was observed in both low- and high-energy images, as well as for objects of varying sizes. Quantitative results showed that the e-Grid method could reduce scatter artifacts by at least 71\% in low-energy images and 75\% in high-energy images.\\
{\bf Significance:} It was demonstrated that the scatter correction method originally applied to DLFPD-based CBCT could also perform well in PCD-CBCT, showing that the e-Grid method has great potential for application in other spectral CBCT imaging systems.\\
\end{abstract}


\maketitle

\section{Introduction}
With the development of computed tomography (CT), the potential benefits of photon-counting detectors (PCDs) have been extensively studied\cite{leng2019photon, rajendran2022first}. With enhanced detective quantum efficiency (DQE) and photon energy resolution\cite{muller2016towards}, PCDs offer superior imaging quality compared to conventional energy-integrating detectors (EIDs), enabling more precise and accurate imaging. This facilitates the application of PCD in interventional radiology(IR)\cite{orth2008c} and image-guided radiation therapy (IGRT)\cite{de2013image} based on CBCT imaging. Several efforts have already been made to integrate large-area PCDs onto C-arm gantries for dual-energy 2D and 3D imaging applications\cite{muller2016towards, ji2021development, ahmad2017assessment, treb2022c}. Depending on the unique energy-discriminating capability, PCDs can filter out low-energy noise and achieve high-quality imaging. However, the use of PCDs in cone-beam imaging is still affected by Compton scatter\cite{ji2021development}, which remains a significant issue, particularly in cone-beam CT (CBCT.) 

Numerous studies have been conducted to address Compton scatter artifacts in CBCT, which can serve as a reference for PCD-based CBCT. Scatter correction methods can be divided into two categories: hardware-assisted corrections and computation-assisted corrections. The principle of hardware-assisted correction methods is to integrate specific hardware into the imaging system to directly reduce scatter or assist in estimating scatter distribution. These methods primarily include anti-scatter grids\cite{neitzel1992grids}, beam blockers\cite{zhu2009scatter, wang2010scatter, niu2011scatter, lee2012scatter}, beam-stop arrays\cite{love1987scatter,ning2004x, siewerdsen2006simple,zhu2005x}, primary modulators\cite{maltz2006cone, zhu2006scatter, ritschl2015robust}, and the air-gap method\cite{neitzel1992grids, siewerdsen2000optimization}. However, it often makes the entire CT imaging system more complex and may increase the patient dose. In addition, computation-assisted scatter correction methods typically obtain scatter distribution through direct image processing and calculation. These methods include Monte Carlo (MC) simulation\cite{colijn2004accelerated, kyriakou2006combining, badal2009accelerating, xu2015practical, sisniega2015high}, kernel-based estimation\cite{hansen1997extraction, ohnesorge1999efficient, maltz2008algorithm, li2008scatter, star2009efficient, sun2010improved}, model-based estimation\cite{meyer2010empirical,zhao2015patient,zhao2016model}, and deep learning approaches\cite{maier2018deep,jiang2019scatter, kida2018cone,kurz2019cbct}. Nevertheless, these methods can hardly achieve a good trade-off between scatter estimation accuracy and computational efficiency.

In this study, a scatter artifact correction method introduced in our previous work \cite{zhang2024cbct} was validated for PCD-CBCT imaging. This method, named e-Grid, was initially applied for scatter correction in dual-layer flat-panel detector (DLFPD) based CBCT imaging. Phantom experiments were conducted to evaluate its effectiveness in PCD-CBCT scatter correction.

\section{Methods and Materials}\label{sec: method}

\subsection{Scatter Correction Process}
\begin{figure}[ht]
   \begin{center}
   \includegraphics[width=1\textwidth]{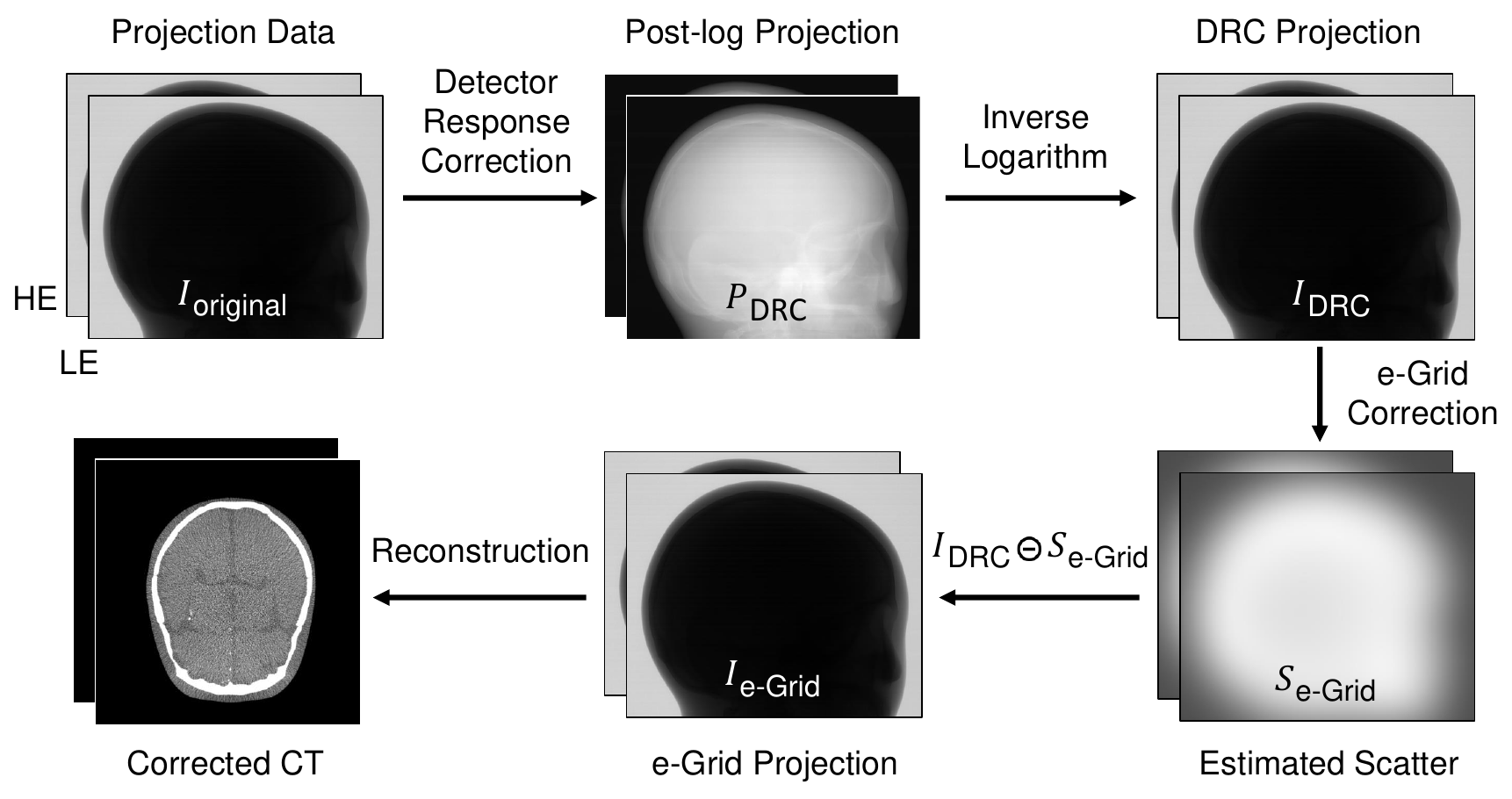}
   \caption{The scatter correction process for PCD-CBCT.} 
   \label{fig:method}
    \end{center}
\end{figure}

The entire scatter correction process is shown in Fig.~\ref{fig:method}. The original projection data is first processed with detector response correction to eliminate band artifacts in the images. These artifacts arise because the PCD consists of many small detector units with different energy responses, which can lead to spectral inconsistencies in the imaging results.

The detector response correction (DRC) method used in this study was proposed by Feng et al.\cite{feng2021experimental}. The main idea behind this method is to calculate the equivalent thickness of aluminum ($T_{\text{Al}}^{\prime}$) and acrylic ($T_{\text{ac}}^{\prime}$). The $T_{\text{Al}}^{\prime}$ is estimated using threshold segmentation, while the $T_{\text{ac}}^{\prime}$ is obtained using the following function:
\begin{equation}
T_{\text{ac}}^{\prime}=c_{1,0}P+ c_{0,1}T_{\text{Al}}^{\prime} +c_{2,0}P^2+ c_{0,2}T_{\text{Al}}^{\prime 2} + c_{1,1}PT_{\text{Al}}^{\prime},\label{eq:Tac}
\end{equation}
where $P$ is the post-log projection data, and the coefficients $c_{\text{i,j}}$ was pre-calibrated using with known thickness aluminum and acrylic. So the post-log projection data after detector response correction can be express as:
Finally, the corrected post-log projection data $P_{\text{DRC}}$ without artifacts was calculated using the following expressing:
\begin{equation}
P_{\text{DRC}}=\mu_{\text{ac}}(E)T_{\text{ac}}^{\prime} + \mu_{\text{Al}}(E)T_{\text{Al}}^{\prime}.\label{eq:PDRC}
\end{equation}
In this study, 10 acrylic sheets each with a thickness of 14.5 mm and 6 aluminum sheets each with a thickness of 0.9 mm were used for the calibration of the $c_{\text{i,j}}$ coefficients. For each thickness, 200 projections were acquired at one angle and averaged for calculation. 

The corrected post-log projection data $P_{\text{DRC}}$ is then logarithmically inverted to obtain the projection data $I_{\text{DRC}}$. In the PCD experiments, two energy thresholds are set, allowing both low-energy data $I_{\text{DRC, LE}}$ (signals between the low-energy threshold and high-energy threshold) and high-energy data $I_{\text{DRC, HE}}$ (signals between the high-energy threshold and tube potential) to be obtained. According to the scatter artifact correction method (e-Grid) introduced in our previous work\cite{zhang2024cbct}, the $I_{\text{DRC, LE}}$ and $I_{\text{DRC, HE}}$ can be expressed in terms of the following linear formulas:

\begin{align}
I_{\text{DRC, LE}}&=I^{\text{p}}_{\text{LE}}+S_{\text{LE}},\label{eq:1}\\
I_{\text{DRC, HE}}&=I^{\text{p}}_{\text{HE}}+S_{\text{HE}},\label{eq:2}
\end{align}
where $I^{\text{p}}_{\text{LE}}$ and $I^{\text{p}}_{\text{HE}}$ denoted the primary low-energy signal intensity and the primary high-energy signal intensity, respectively, $S_{\text{LE}}$ and $S_{\text{HE}}$ denoted the scattered low-energy signal and the scattered high-energy signal, respectively. To retrieve the scattered signals, the following approximations were assumed,

\begin{align}
I^{\text{p}}_{\text{LE}}&=f_{\text{p}}(I^{\text{p}}_{\text{HE}})\approx\alpha^{\text{p}}_{1}I^{\text{p}}_{\text{HE}}+\alpha^{\text{p}}_{0},\label{eq:3}\\
S_{\text{LE}}&=f_{\text{s}}(S_{\text{HE}})\approx \alpha^{\text{s}}_{1}S_{\text{HE}}+\alpha^{\text{s}}_{0},\label{eq:4}
\end{align}
where function $f_{\text{p}}$ and $f_{\text{s}}$ were assumed to map the high-energy signals onto the low-energy signals\cite{brunner2011prior}. Moreover, functions $f_{\text{p}}$ and $f_{\text{s}}$ were approximated by linear expansions with pre-calibrated first-order coefficients $\alpha^{\text{p}}_{1}$, $\alpha^{\text{s}}_{1}$ and zero-order coefficients $\alpha^{\text{p}}_{0}$, $\alpha^{\text{s}}_{0}$. Eventually, the high-energy scatter signal $S_{\text{HE}}$  and low-energy scatter signal $S_{\text{LE}}$ can be obtained through the above four equations (Eq.~(\ref{eq:1}) to Eq.~(\ref{eq:4})). Then, an additional process is applied to remove the high-frequency components in the calculated scatter maps to obtain the final scatter estimation $S_{\text{e-Grid}}$. The details of scatter calculation can be found in our previous work\cite{zhang2024cbct}. Then, the scatter estimation $S_{\text{e-Grid}}$ is subtracted from the projection data $I_{\text{DRC}}$ to obtain the scatter-free data $I_{\text{e-Grid}}$, and the filtered backprojection (FBP) algorithm is applied to reconstruct the final corrected CT images.

\subsection{Phantom Experiments}
\begin{figure}[t]
   \begin{center}
   \includegraphics[width=0.8\textwidth]{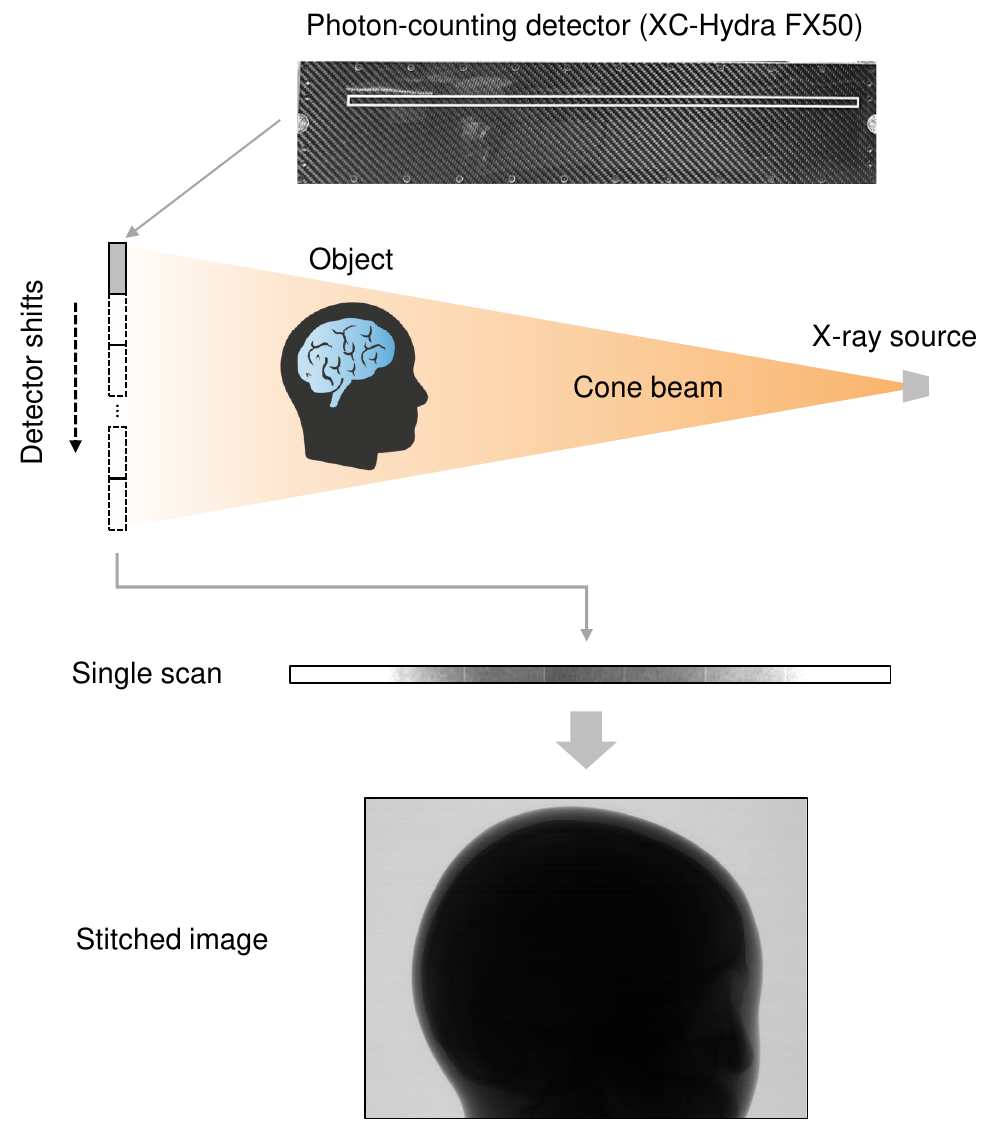}
   \caption{Illustration of a PCD-CBCT imaging system.}
   \label{fig:schematic}
    \end{center}
\end{figure}
In this study, a Cadmium Zinc Telluride (CZT)-based photon-counting detector (XC-Hydra FX50, Direct Conversion AB, Sweden) with $5120 \times 64$ elements, each measuring 0.1~mm $\times$ 0.1~mm, was used to perform PCD-CBCT imaging. To achieve cone beam CT imaging, a shift-scan method was applied to capture a cone beam field of view (FOV), with projections stitched together to generate the phantom imaging data. As shown in Fig.~\ref{fig:schematic}, during the shift scan, the detector was shifted 5 mm from the bottom to the top of the object with each scan, while the X-ray source remained in cone beam scan mode. Two phantoms were used in the experiments: a 16 cm diameter angiographic head phantom (Model: 41309-300, Kyoto Kagaku, Japan) and a 28 cm diameter abdominal phantom (Model: 057A, CIRS, USA). The number of detector shifts was 47 for the head phantom and 43 for the abdominal phantom.

The source-to-detector distance (SDD) was fixed at 1100 mm and the source-to-object distance (SOD) was fixed at 850 mm. The 450 projections were acquired over a rotation angle of 360 degrees. A tube voltage of 120 kVp was applied, with a fixed 1.5 mm aluminum beam filter. Additionally, a 0.4 mm copper filter was used for the head phantom, and a 0.8 mm copper filter for the abdominal phantom. More detailed experimental settings were listed in Table~\ref{tab:parameter}.

The weighted CT Dose Index ($\rm{CTDI_w}$) was experimentally determined using a 16 cm CTDI phantom and a 100 mm pencil ionization chamber (X2 CT Sensor, RaySafe, Sweden). The measured $\rm{CTDI_w}$ was 34.68 mGy for the head phantom and 20.51 mGy for the abdominal phantom. It should be noted that the $\rm{CTDI_w}$ for the 28 cm abdominal phantom was also measured using the 16 cm CTDI phantom.

\begin{table}[!htb]
\begin{center}
\caption{The parameters used in physical experiments. 
\vspace*{2ex}
}
\renewcommand\arraystretch{0.9}
\resizebox{0.65\columnwidth}{!}{
\begin{tabular} {p{5.5cm}p{5cm}}
\toprule[1pt]
\hline
  Parameter&Value\\
\cline{1-2}
Tube potential (kVp) &120\\
Tube current (mA) &12.5 \\
Beam filtration: Al (mm) &1.5 \\
Beam filtration: Cu (mm) &Head phantom: 0.4;  Abdominal phantom: 0.8\\
Total views&450\\
Frame rate (fps) &10\\
SOD (mm)&850\\
SDD (mm)&1100\\

Detector array&$5120\times64$\\
Pixel size ($\rm{mm}^2$)&$0.1\times0.1$\\
Low Energy threshold (keV)  &20\\
High Energy threshold (keV)  &65\\
\hline
\bottomrule[1pt]
\end{tabular}}
\label{tab:parameter}
\end{center}
\end{table}

In addition to the PCD-shifted CBCT imaging, PCD fan beam CT imaging was also performed as a reference. Compared to the cone beam (with a 23 cm beam width on the detector plane), the fan beam (with a 0.5 cm beam width on the detector plane) was assumed to contain negligible scatter signals. Note that all the experimental results were processed with detector response correction.

\subsection{Quantitative metric}
To assess the effectiveness of the correction, the image non-uniformity (NU) indices were calculated, defined as follows:
\begin{equation}
\text{NU}=\left|\frac{{\bar{\mu}_{\text{c}}}-{\bar{\mu}_{\text{e}}}}{{\bar{\mu}_{\text{e}}}}\right|\times 100\%,\label{eq:nu}
\end{equation}
where ${\bar{\mu}_{\text{c}}}$ and ${\bar{\mu}_{\text{e}}}$ represent the mean values of the regions of interest (ROIs) selected at the center and the edge of the reconstructed CBCT images, respectively.

In addition, the signal-to-noise ratio (SNR) of the CT images was compared before and after scatter correction. The SNR of the CT images was defined as:
\begin{equation}
\text{SNR}=\frac{{\bar{\mu}}}{{\sigma}},\label{eq:snr}
\end{equation}
where $\bar{\mu}$ denoted the mean value of the selected ROI on the reconstructed CT images, and $\sigma$ denoted the corresponding standard deviation.

\section{Results}\label{sec: results}
\subsection{Head phantom experimental results}
\begin{figure}[!htb]
   \begin{center}
   \includegraphics[width=1\textwidth]{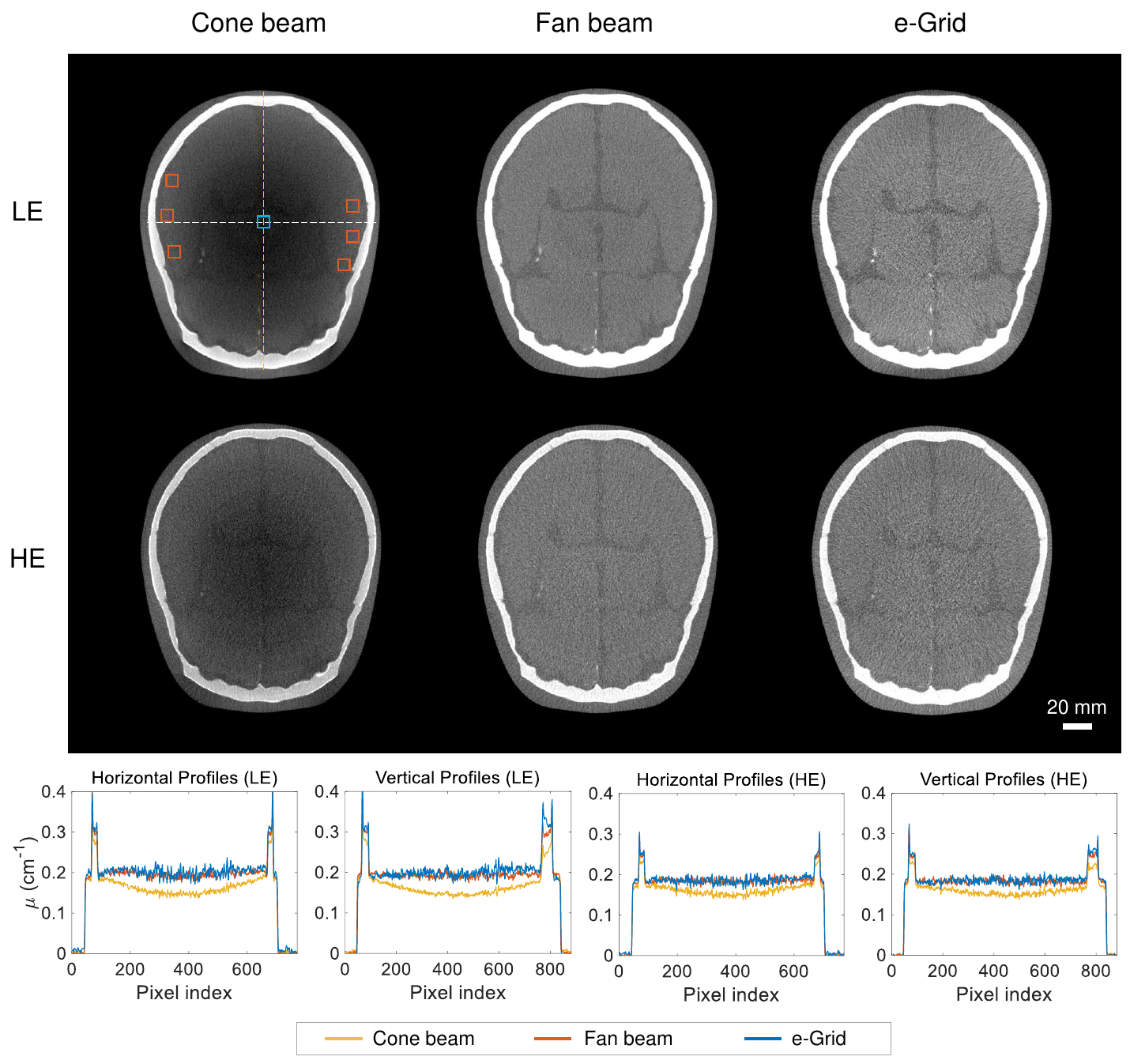}
   \caption{PCD-CBCT imaging results of the head phantom. The profiles along the horizontal and vertical directions were compared and plotted at the bottom. The display window was $\left[0.13, 0.28\right]$ $\rm{cm}^{-1}$ for low-energy CT images, and $\left[0.13, 0.25\right]$ $\rm{cm}^{-1}$ for high-energy CT images. The scale bar denoted 20 mm.}
   \label{fig:KYOTOCT}
    \end{center}
\end{figure}

\begin{figure}[!htb]
   \begin{center}
   \includegraphics[width=1\textwidth]{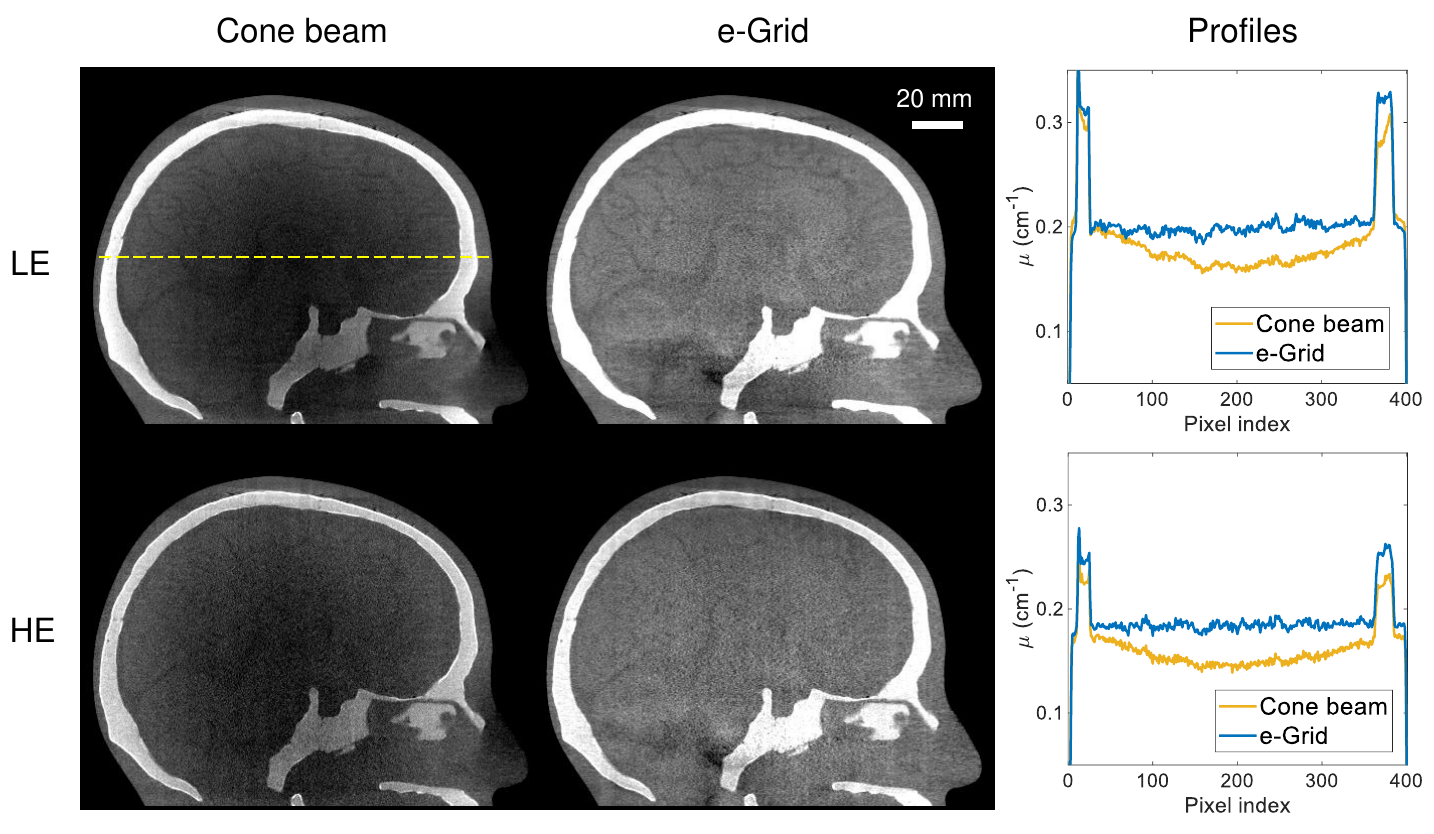}
   \caption{PCD-CBCT imaging results of the head phantom on the sagittal view plane. The profiles along the horizontal directions were compared and plotted on the right-hand side. The display window was $\left[0.13, 0.28\right]$ $\rm{cm}^{-1}$ for low-energy CT images, and $\left[0.13, 0.25\right]$ $\rm{cm}^{-1}$ for high-energy CT images. The scale bar denoted 20 mm.}
   \label{fig:KYOTOCT_Sagittal}
    \end{center}
\end{figure}

The PCD-CBCT experimental results for the head phantom are shown in Fig.~\ref{fig:KYOTOCT}. It should be noted that the results for cone beam and fan beam imaging were obtained with only detector response correction, while the e-Grid results were obtained with both detector response correction and e-Grid scatter correction. It could be observed that the head phantom exhibited noticeable cupping artifacts in the cone beam imaging results, whereas such artifacts were hardly visible in the fan beam imaging results. However, after applying the e-Grid scatter correction, as shown in the third column of Fig.~\ref{fig:KYOTOCT}, most of the cupping artifacts were eliminated. The brain tissue structure in the middle of the head phantom was clearly visible after scatter correction, and the overall image quality was improved. These observations were consistent in both low-energy and high-energy imaging results. Additionally, from the profile plot at the bottom, it could be observed that the curve of the cone beam exhibited a concave shape in the middle. Nevertheless, the curve became flatter with the e-Grid scatter correction, resulting in a more uniform signal distribution that closely resembled the fan beam results. It was consistent in both the horizontal and vertical directions. These profile trends aligned with the visual observations above and indicated that the e-Grid method effectively suppressed Compton scatter in PCD-CBCT imaging, thereby improving image quality.

The PCD-CBCT imaging results of the head phantom on the sagittal view plane are shown in Fig.~\ref{fig:KYOTOCT_Sagittal}. It could be observed that Compton scatter was also mitigated through the e-Grid method in the sagittal view. The inner tissue became more visible, and the overall image uniformity was improved.

\subsection{Abdominal phantom experimental results}
\begin{figure}[!htb]
   \begin{center}
   \includegraphics[width=1\textwidth]{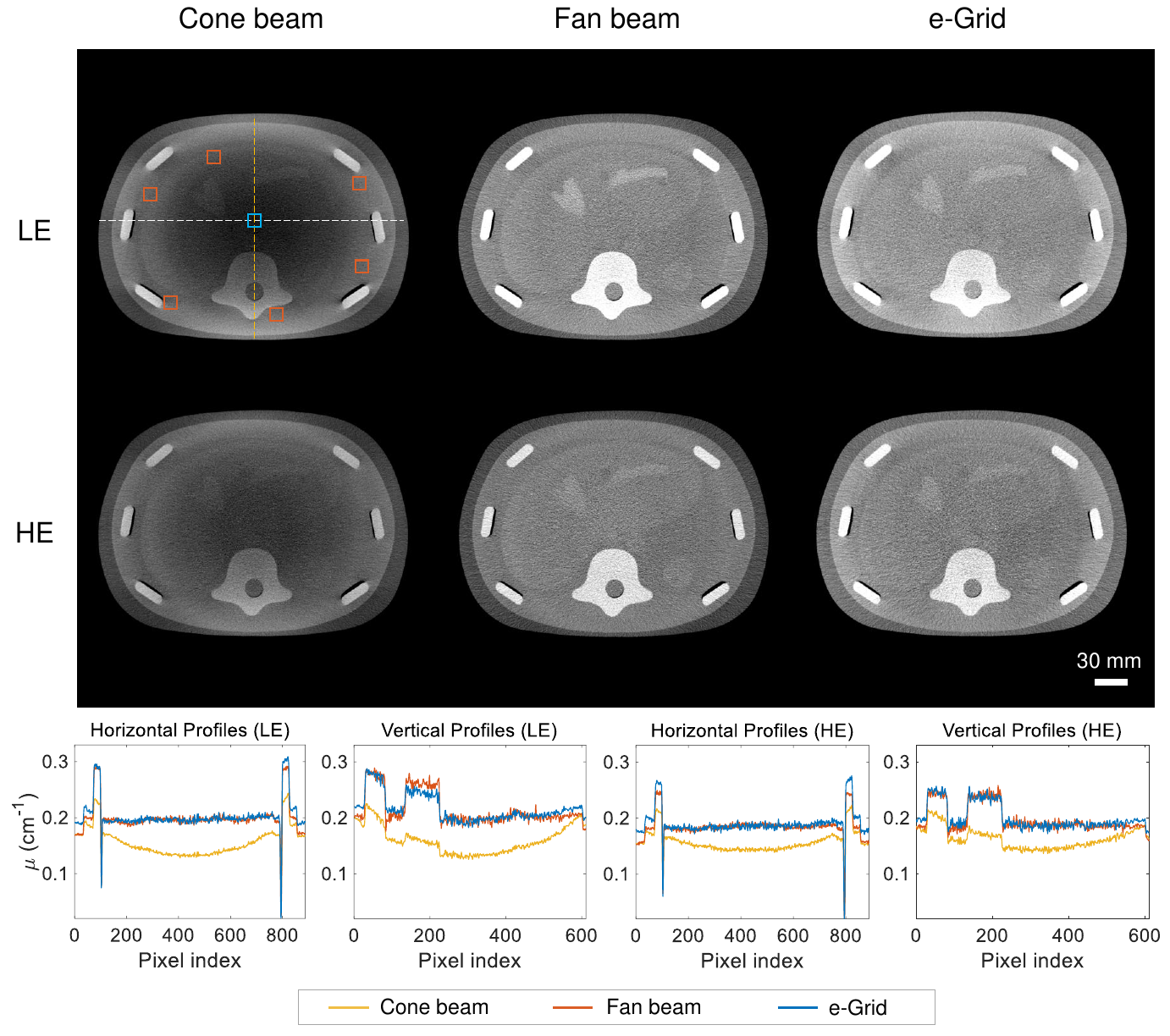}
   \caption{PCD-CBCT imaging results of the abdominal phantom. The profiles along the horizontal and vertical directions were compared and plotted at the bottom. The display window was $\left[0.12, 0.27\right]$ $\rm{cm}^{-1}$ for low-energy CT images, and $\left[0.11, 0.26\right]$ $\rm{cm}^{-1}$ for high-energy CT images. The scale bar denoted 30 mm.}
   \label{fig:AbdomenCT}
    \end{center}
\end{figure}
\begin{figure}[!htb]
   \begin{center}
   \includegraphics[width=1\textwidth]{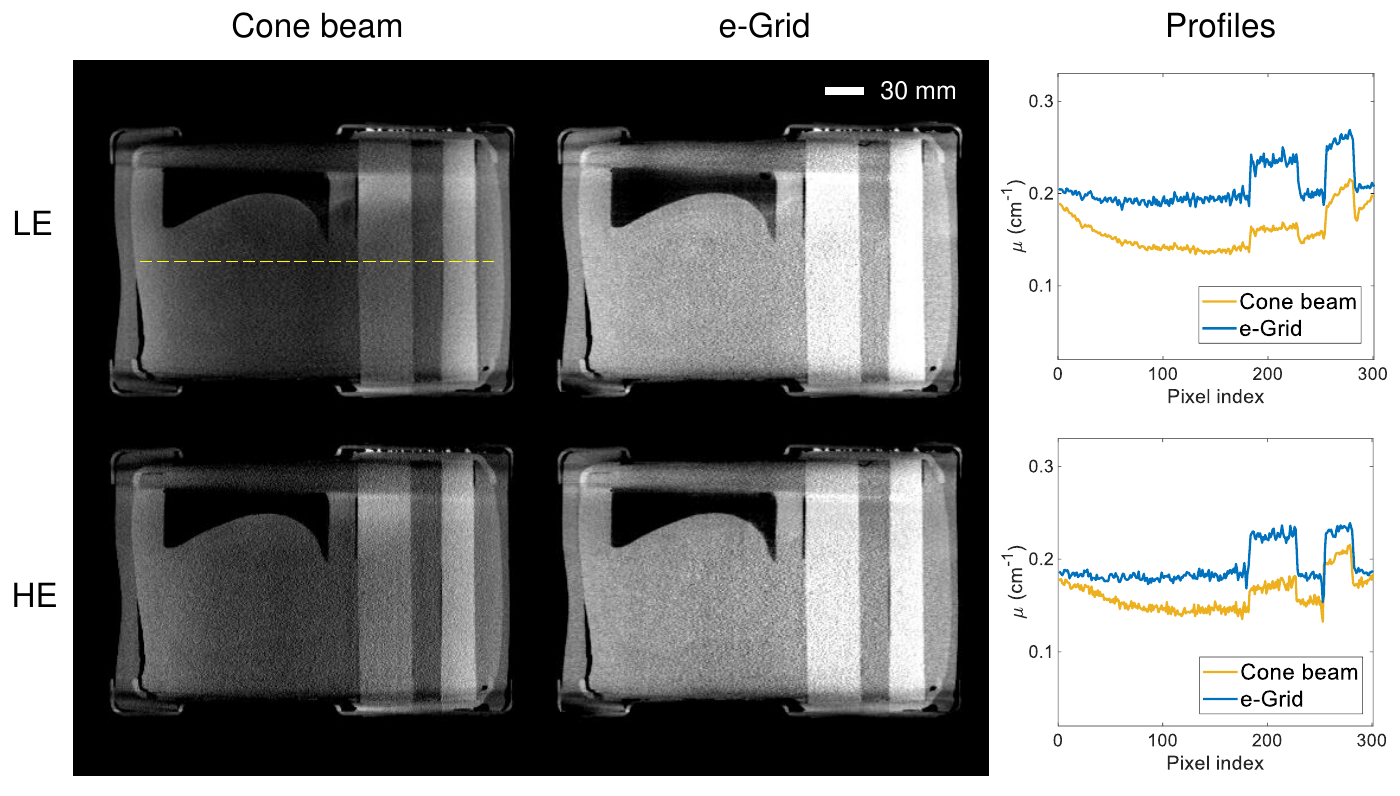}
   \caption{PCD-CBCT imaging results of the abdominal phantom on the sagittal view plane. The profiles along the horizontal directions were compared and plotted on the right-hand side. The display window was $\left[0.12, 0.27\right]$ $\rm{cm}^{-1}$ for low-energy CT images, and $\left[0.11, 0.26\right]$ $\rm{cm}^{-1}$ for high-energy CT images. The scale bar denoted 30 mm.}
   \label{fig:AbdomenCT_Sagittal}
    \end{center}
\end{figure}

The PCD-CBCT experimental results for the abdominal phantom were depicted in Fig.~\ref{fig:AbdomenCT}. Similar to the results for the head phantom, numerous scatter artifacts remained in the PCD-CBCT imaging results of the abdominal phantom, which obscured the internal imaging information. Though, unlike the head phantom, the abdominal phantom had a larger diameter of 28 cm, when the e-Grid method was applied to correct PCD-CBCT images, the shading artifacts were removed, resulting in high-quality imaging. It can be seen that, in relation to the fan beam results, which were considered scatter-free imaging, the e-Grid method showed comparable imaging performance. And this high imaging performance was present in both low-energy and high-energy images. Meanwhile, the horizontal and vertical direction line profiles showed that the e-Grid corrected results significantly improved the uniformity of the CT images, demonstrating the effectiveness of this method in PCD-CBCT imaging for large objects from another perspective.

The PCD-CBCT imaging results of the abdominal phantom in the sagittal plane are shown in Fig.~\ref{fig:AbdomenCT_Sagittal}. Consistent with the observation in Fig.~\ref{fig:AbdomenCT}, the e-Grid method effectively reduced cupping artifacts in cone beam imaging results and generated high-quality CT images. The line profile results also supported these findings.

\subsection{Quantitative measurement results}
\begin{figure}[!htb]
   \begin{center}
   \includegraphics[width=1\textwidth]{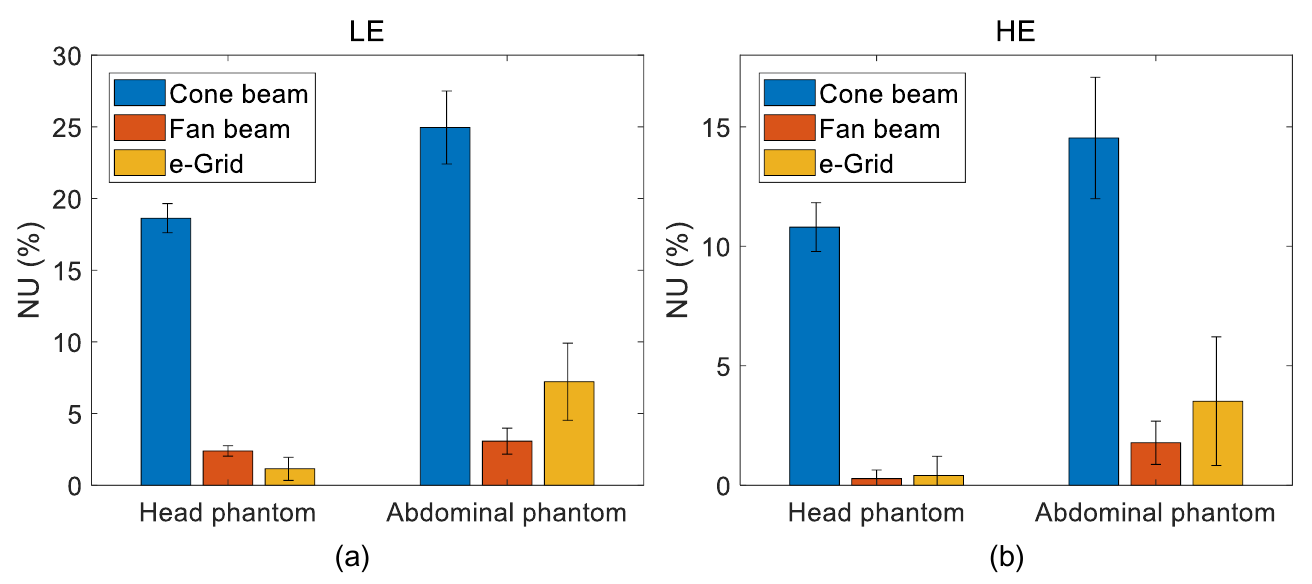}
   \caption{The measured non-uniformity (NU) indices on the (a) low-energy (LE) CT images and (b) high-energy (HE) CT images for the head phantom and abdominal phantom, respectively.}
   \label{fig:NU}
    \end{center}
\end{figure}

The image uniformity of the imaging results for the head phantom and abdominal phantom was measured, with the statistical results shown in Fig.~\ref{fig:NU}. The selected ROIs for NU calculation for the head phantom and abdominal phantom are shown the first image of Fig.~\ref{fig:KYOTOCT} and Fig.~\ref{fig:AbdomenCT}, respectively. Therein, the blue-box area represented the central ROI, while the six orange-box areas represented the edge ROIs. The variations in phantom size and ROI positions resulted in slight differences in the NU values between the head phantom and the abdominal phantom. 

From the histogram, it can be observed that due to the presence of scatter artifacts in the uncorrected PCD-CBCT images, there is a difference in values between the central ROI and the peripheral ROIs. As a result, the NU value of the original cone beam imaging results is relatively high. However, when the e-Grid method was used to reduce the cupping artifacts in the original PCD-CBCT images, the NU measurement values showed a significant decrease. Compared to the uncorrected cone beam results, the NU value measured on the e-Grid processed low-energy CT images was reduced by over 93\% for the head phantom and 71\% for the abdominal phantom. Additionally, for high-energy imaging, the NU value measured on the e-Grid results was reduced by over 88\% for the head phantom and 75\% for the abdominal phantom.

Additionly, the measured SNR results are shown in Table~\ref{tab:snr}. The selected ROIs for SNR calculation, shown the first image of Fig.~\ref{fig:KYOTOCT} and Fig.~\ref{fig:AbdomenCT}, are consistent with those used for NU calculation. It can be observed that the SNR of the e-Grid processed low-energy CT images decreased by 38\% for the head phantom and 17\% for the abdominal phantom. For high-energy CT images, the SNR decreased by approximately 19\% for the head phantom and 20\% for the abdominal phantom. This reduction was primarily attributed to the removal of scattered X-rays\cite{wang2010scatter, niu2011scatter}.

\begin{table}[!htb]
\begin{center}
\caption{The measured SNR values for low-energy and high-energy CT images.
\vspace*{2ex}
}
\renewcommand\arraystretch{1.3}
\resizebox{0.9\columnwidth}{!}{
\begin{tabular} {l@{\hspace{0.5cm}}c@{\hspace{0.5cm}}c@{\hspace{0.5cm}}c@{\hspace{0.5cm}}c}
\toprule[1pt]
\hline
&Cone beam (LE)&e-Grid (LE)&Cone beam (HE)&e-Grid (HE)\\
\cline{1-5}
Head phantom& 34.28$\pm$5.68 & 21.16$\pm$3.93 & 26.97$\pm$3.81 & 22.38$\pm$3.20 \\
Abdominal phantom& 25.13$\pm$4.56 & 20.43$\pm$3.76 & 21.83$\pm$2.80 & 17.40$\pm$ 2.55 \\
\hline
\bottomrule[1pt]
\end{tabular}}
\label{tab:snr}
\end{center}
\end{table}

\section{Discussions and conclusion}\label{sec: diss}
In this study, the effectiveness of a scatter correction method originally developed for DLFPD-based CBCT imaging was validated for scatter suppression in PCD-CBCT. The results of the phantom experiments showed that the method, named e-Grid, achieved good performance in PCD-CBCT scatter correction, significantly improving the image quality degraded by scatter artifacts. This remarkable imaging performance was observed in both low-energy and high-energy images, as well as across varying sizes of imaged objects. Meanwhile, the quantitative measurements also demonstrated that the e-Grid method performed well in PCD-CBCT, effectively eliminating the cupping artifacts and improving overall image uniformity. In generally, the e-Grid method can reduce scatter artifacts by at least 71\% in low-energy images and 75\% in high-energy images.

These results indicated that the e-Grid method, originally developed for DLFPD-based CBCT imaging, could be extended to scatter correction for other spectral CBCT systems. This may be attributed to the method’s reliance on energy modulation during imaging process, which allows it to be effective in CBCT systems with energy discrimination capabilities. Therefore, for other similar spectral CBCT systems, such as fast kV switching and dual-source systems, it is anticipated that e-Grid could also be applied for scatter correction, enhancing the imaging quality of these systems.

In conclusion, this study demonstrates that the e-Grid scatter correction method can be applied to PCD-CBCT systems, effectively addressing the scatter artifacts present in the imaging results. It also indicates that this method holds strong potential for use in other spectral CBCT systems.

\section*{Funding}
This work was supported in part by the National Natural Science Foundation of China (12027812, 62201560, U23A20284, 12305349, 12235006), Shenzhen Science and Technology Program (JSGGKQTD20210831174329010), Youth Innovation Promotion Association of Chinese Academy of Sciences (2021362).

\bibliography{./Bibliography_Paper}
\bibliographystyle{./medphy}
\end{document}